\documentstyle[epsfig,cite]{article}    % Specifies the document style.
 
\large
\newcommand{\Title}[1]{\begin{center}{\Large\bf #1}
\end{center}\vskip 0.1in}
\newcommand{\Name}[1] {\begin{center}{\large    #1}
\end{center}\vskip 0.1in}

\def\Thebibliography#1{\section*{REFERENCES}\list
 {[\arabic{enumi}]}{\settowidth\labelwidth{[#1]}\leftmargin\labelwidth
 \advance\leftmargin\labelsep
 \usecounter{enumi}}
 \def\newblock{\hskip .11em plus .33em minus .07em}
 \sloppy\clubpenalty4000\widowpenalty4000
 \sfcode`\.=1000\relax}

\newcommand{\FigureAng}[4]{
  \mbox{\epsfig{file=#1,width=#2,height=#3,angle=#4}}
  }
 \newcommand{\be}{\begin{equation}}
\newcommand{\ee}{\end{equation}}
\newcommand{\ba}{\begin{array}{c}}
\newcommand{\ea}{\end{array}}
\newcommand{\beqn}{\begin{eqnarray}}
\newcommand{\eeqn}{\end{eqnarray}}

\def\beq{\begin{equation}}
\def\eeq{\end{equation}}
\newcommand{\bea}{\begin{eqnarray}}
\newcommand{\eea}{\end{eqnarray}}
\newcommand{\nn}{\nonumber}
\def\simleq{\; \raise0.3ex\hbox{$<$\kern-0.75em
      \raise-1.1ex\hbox{$\sim$}}\; }
\def\simgeq{\; \raise0.3ex\hbox{$>$\kern-0.75em
      \raise-1.1ex\hbox{$\sim$}}\; }
\def\noi{\noindent}

\def\R{ {\rm R \kern -.31cm I \kern .15cm}}
\def\C{ {\rm C \kern -.15cm \vrule width.5pt
\kern .12cm}}
\def\Z{ {\rm Z \kern -.27cm \angle \kern .02cm}}
\def\N{ {\rm N \kern -.26cm \vrule width.4pt \kern .10cm}}
\def\1{{\rm 1\mskip-4.5mu l} }
 
\begin{document}
\large
%\tableofcontents
% \listoftables
% \listoffigures
 
\Title{Inclusive decay of $B$ mesons into $D_s$ or $D^*_s$}
 
\Name{{\bf R. Aleksan, M. Zito} \\
Commissariat \`a l'Energie Atomique, Saclay \\
DSM/DAPNIA/SPP \\
91191 Gif-sur-Yvette Cedex, France}

\Name{{\bf A. Le Yaouanc, L. Oliver, O. P\`ene and J.-C. Raynal} \\
Laboratoire de Physique Th\'eorique \footnote{
Unit\'e Mixte de Recherche CNRS - UMR 8627} \\
Universit\'e de Paris XI, B\^atiment 210\\
91405 Orsay Cedex, France }
\vskip 1 truecm
\centerline{\bf Abstract} \par \vskip 3 truemm

We compute the inclusive decay rates $b \to D_s^-(D^{*-}_s) c$ including lowest
order QCD corrections on the quark legs, and compare with existing data. Unlike the
short distance QCD corrections, that are of higher order, these corrections are of
order $\alpha_s$. In the on-shell renormalization scheme and for $\alpha_s(m_b) \cong 0.2$ we
find a correction of  $- 10$~$\%$ to the inclusive rate computed using factorization. This gives a
total rate $BR(b \to D_s^-(D^{*-}_s)c) \cong 8 \ \%$ consistent within $1\sigma$ with the
measured value $BR(B \to D_s^{\pm}X) = (10.0 \pm 2.5)\ \%$. The general formulae given here include
the case of vanishing mass for the final quark $b \to D_s^-(D_s^{*-})u$. The radiative
correction to this rate is $- 17$~$\%$. We show in another place that this process can be
useful for the measurement of the CKM matrix element $V_{ub}$. We also give the
renormalized vertex at the interesting values $q^2 = 0$ and $q^2 = q_{max}^2$, and
compare with existing literature. 

\vskip 1 truecm 

\noi
LPT Orsay 99-36 \par 
\noi DAPNIA/SPP 99-19 \par
\noi May 1999 \par \newpage
\pagestyle{plain}
\baselineskip = 26 pt 
\section{Introduction} \hspace{\parindent} The inclusive rate of a $\bar{B}$ meson
decaying into a $D_s^-$ meson is obtained using the spectator quark model shown in
Fig.~1~:
\be
\label{1e}
\Gamma (\bar{B} \to D_s^- X) \cong \Gamma (b \to D_s^-c) + \Gamma (b \to D_s^{*-}c) \quad
.  \ee  

\noi A main point in writing the approximation (1) is that, as we point out in ref.
\cite{1r}, the excited states $D_s^{**}$ do not lead to $D_s$ since their dominant
decays are $D_s^{**} \to D^{(*)}K$. Moreover, the quark $c$ dominates the inclusive
rate, the quark $u$ being CKM suppressed. Of course, there are also other mechanisms
for producing the $D_s$ of the right sign, namely through annihilation and exchange
diagrams. However, as we discuss in detail in ref. \cite{1r}, these mechanisms are
suppressed. In this paper we will concentrate on the mechanism of Fig.~1, and compute
the $O(\alpha_s)$ radiative corrections to it. \par
%%%%%%%%%%%%%%%%%%% Figure %%%%%%%%%%%%%%%%%%%%%%%%%%
\setlength{\unitlength}{0.7mm}
\begin{figure*}[htb]
\vfill
\begin{picture}(280,80)(-50,-80)
%\begin{center}
\FigureAng{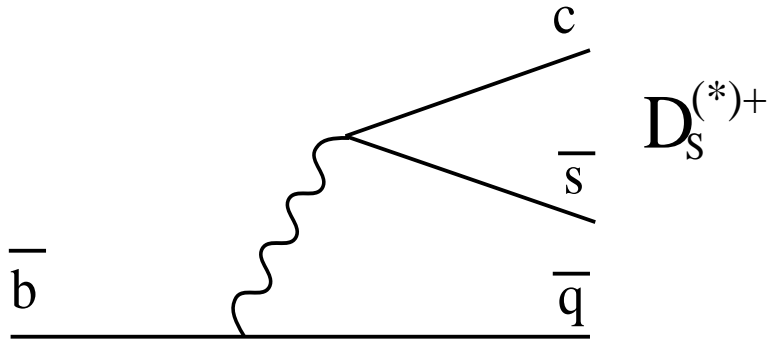}{5.0cm}{10.0cm}{-90}
%\FigureAng{babardwg.eps}{9.0cm}{12.0cm}{90}
%\mbox{\epsfig{file=/home/usr201/mnt/aleksan/btods_fig1.eps}}
%\end{center}
\end{picture}
\vfill
\caption{Spectator diagram for the decay 
$\overline{b}\rightarrow D^+_s \overline{q}$. }
\label{fig:diagsp}
\end{figure*}
%%%%%%%%%%%%%%%%%%%% Figure Unitarity Triangle %%%%%%%%%%%%%

~From the effective weak Hamiltonian, using factorization and the spectator quark model,
a straightforward calculation gives the rates
\be 
\label{2e}
\Gamma^{(0)}(b \to D_s^-q) = {G^2 \over 16 \pi} \ |V_{qb}V_{cs}^*|^2 \ a^2_1 \ f_{D_s}^2 \
m_b^3 \sqrt{\lambda (1, r^2, \xi)} \left [ (1 - r^2)^2 - \xi (1 + r^2) \right ]
 \ee  

\be 
\label{3e}
\Gamma^{(0)}(b \to D_s^{*-}q) = {G^2 \over 16 \pi} \ |V_{qb}V_{cs}^*|^2 \ a^2_1 \
f_{D_s^*}^2 \ m_b^3 \sqrt{\lambda (1, r^2, \xi)} \left [ (1 - r^2)^2 + \xi (1 + r^2  - 2
\xi) \right ]
 \ee 

\noi where $a_1$ is the QCD short distance factor
\be
\label{4e}
a_1 = C_2 + {C_1 \over N_c} = {c_+ + c_- \over 2} + {c_+ - c_- \over 2 N_c} \quad .
\ee

\noi Empirically, from exclusive decays, one finds $a_1$ consistent with $|a_1| \cong
1$. For further convenience we have adopted the notation
\be
\label{5e}
r = {m_q \over m_b} \qquad \xi = {q^2 \over m_b^2}
\ee

\noi where $m_q$ is the final quark mass $m_q = m_c$ or $m_u$, $q^2 = m_{D_s}^2$, or
$m_{D_s^*}^2$, and $\lambda (a,b,c) = a^2 + b^2 + c^2 - 2ab - 2bc - 2ca$. \par

To compute the order of magnitude of the expected branching fractions, let us tentatively use the
following numerical values for the quark masses (see below for a detailed discussion of these parameters) 
\be
\label{6e}
m_b = 5.0 \ \hbox{GeV} \quad , \qquad m_c = 1.6 \ \hbox{GeV} 
 \ee

\noi and $m_u \cong 0$ and the decay constants \cite{3r}
\be
\label{7e}
f_{D_s} = 230 \ \hbox{MeV} \qquad f_{D_s^*} = 280 \ \hbox{MeV} \quad .
\ee

\noi Using $\tau_B = 1.6$ ps and $|V_{cb}| = 0.04$, one obtains
\be
\label{8e}
BR^{(0)} (b \to D_s^-c) \cong 3.2 \ \% \qquad BR^{(0)} (b \to D_s^{*-}c) \cong 6.8\ \% \quad .
\ee 

\noi where the superindex $(0)$ means using the factorization results (\ref{2e}), (\ref{3e}).
Within the assumption (\ref{1e}), this yields \be
\label{9e}
BR (\bar{B} \to D_s^-  X) \cong BR^{(0)} (b \to
D_s^-c) + BR^{(0)} (b \to D_s^*c) \cong 10 \ \% \ee

\noi and, for completeness, with ${|V_{ub}| \over |V_{cb}|} = 0.08$, let us give the $b
\to u$ branching ratios~:
\be
\label{10e}
BR^{(0)} (b \to D_s^-u) \cong 2.6 \times 10^{-4} \qquad BR^{(0)} (b \to D_s^{*-}u) \cong 4.9
\times 10^{-4} \quad . \ee

\noi The naive prediction (\ref{9e}) is consistent with the measured value 
\cite{4bis}
\be
\label{11e}
BR(B \to D_s^{\pm}  X) = (10.0 \pm 2.5) \ \% \quad .
\ee

The aim of this paper is to compute the lowest order QCD corrections to the process
depicted in Fig.~1, i.e. to investigate how stable is the naive result (\ref{9e})
relatively to QCD radiative corrections. It is important to emphasize that the calculation of the radiative corrections will ask
in particular for a detailed discussion of the quark masses, that we perform in Section 6. \par

%%%%%%%%%%%%%%%%%%% Figure %%%%%%%%%%%%%%%%%%%%%%%%%% 
\setlength{\unitlength}{0.7mm} \begin{figure*}[htb] \vfill
\begin{picture}(280,140)(-35,-140) %\begin{center}
\FigureAng{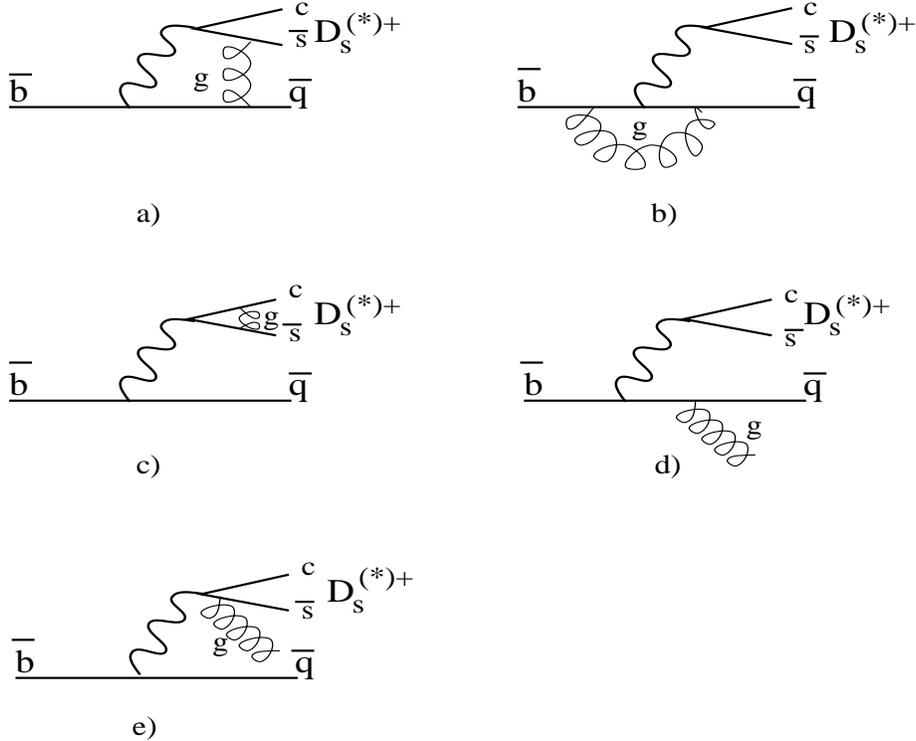}{10.0cm}{12.0cm}{-90}
%\FigureAng{babardwg.eps}{9.0cm}{12.0cm}{90}
%\mbox{\epsfig{file=/home/usr201/mnt/aleksan/btods_fig1.eps}} %\end{center}
\end{picture}
\vfill
\caption{Diagrams leading to the production of $D_s$ mesons. }
\label{fig:diagbkg}
\end{figure*}
%%%%%%%%%%%%%%%%%%%% Figure Unitarity Triangle %%%%%%%%%%%%%
Let us first notice that the leading logarithm
corrections cancel at order $\alpha_s$. Indeed, from the well known expressions at
leading order $c_{\pm}(\mu ) = \left [ {\alpha_s(\mu) \over \alpha_s (m_W)} \right
]^{d_{\pm}}$ where $d_+ = -{6 \over 23}$, $d_- = {12 \over 23}$, one finds for the
expression (\ref{4e}) $a_1 = 1 + O(\alpha_s^2)$. The correction is at most of second
order, and this explains why the combination of short distance coefficients $a_1$ is
predicted by the theory to be close to one. This agrees with the measurements
on class I exclusive decays \cite{4r}, that yield an empirical $|a_1| = 1.00 \pm 0.06$
\cite{5r} when one uses SVZ factorization \cite{6r} to calculate the matrix elements. \par

Figure 2 depicts the lower order QCD corrections. The contribution of Fig.~2a corresponds to
gluons going from the lower to the upper vertex, they are UV divergent and contribute to the
anomalous dimension. Figs.~2b,c are UV convergent if one adds the corresponding self-energy
diagrams on the quark legs~: this corresponds to the well-known fact that currents have no
anomalous dimension. The fact that, as we have seen, the contribution of Fig.~2a vanishes at
$O(\alpha_s)$ follows from the color neutrality of the $D_s^{(*)}$. Indeed, the $W$ produces a
$q\bar{q}$ in a color singlet. If the latter emits only one gluon which goes to the other vertex,
and does not reabsorb any gluons, it becomes a color octet and cannot end into a $D_s^{(*)}$. Of
course, this vanishing at $O(\alpha_s)$ holds in general, not only for the leading
$\alpha_s \log \left ( {m_W^2 \over \mu^2} \right )$ term in the combination $a_1$
(\ref{4e}). \par

Since $O(\alpha_s)$ leading log QCD corrections vanish, we will have to consider
$O(\alpha_s)$ corrections not containing the coefficient $\log \left ( {m_W^2 \over
\mu^2}\right)$, i.e. we will have to consider also the UV finite radiative
corrections affecting {\it separately} the lower and upper quark legs. These corrections
are IR divergent, and have to be combined with the Bremsstrahlung graphs on the same leg
(Figs.~2d,e) to obtain the IR convergent physical result. At the upper vertex, however,
all the gluon corrections of the type 2c, e (and the corresponding self-energies) have to
be included in the vacuum to $D_s^{(*)}$ matrix element represented by the decay
constants $f_{D_s}$ or $f_{D_s^*}$. Therefore, if we use a consistent evaluation of the
latter, we will not have to bother about gluons at the upper vertex. \par

In the following, we present an estimation of the $O(\alpha_s)$ radiative corrections at
the lower vertex ($b$-$q$ quark line), for arbitrary $q$ quark mass and $q^2$ carried by
the current. Of course, the interesting cases are $m_q = m_c$ or
$m_u$ and $q^2 = m_{D_s}^2$ or $m_{D^*_s}^2$. Calculations of radiative corrections at the order aimed in this paper have been made in
the past by Guberina, Peccei and R\"uckl \cite{7r}, that computed the $O(\alpha_s)$
corrections to the total rate $b \to q_1q_2\bar{q}_3$ in the case of {\it vanishing
masses} for the final quarks. The result is the sum of two contributions~: the QCD
translation of the Berman, Kinoshita and Sirlin calculation of QED corrections
to $\mu$ decay and the $O(\alpha_s)$ correction to the ratio $R$ in $e^+e^-$
annihilation. Guberina et al. used, as we do in this paper, naive dimensional
regularization (NDR) \cite{10r} for both UV and IR divergences. From their intermediate results
for what is now for us the lower current in Fig.~2, we can compute, saturating with the tensors
$f_{D_s}^2 q^{\mu} q^{\nu}$ (for the $D_s$) or $f_{D^*_s}^2(-g^{\mu \nu} q^2 + q^{\mu}q^{\nu})$
(for the sum over the polarizations of the $D_s^*$), with $q^2 = m_{D_s}^2$ or $m_{D^*_s}^2$,
the total rates $\Gamma (b \to D_s^-u)$ and $\Gamma (b \to D_s^*u)$, since the $u$ quark
mass is approximately massless. We have done it, but we need to make the general
calculation for a final massive quark $q$. In the limit $m_q \to 0$ we recover
the result that we have obtained from the intermediate stages worked out by Guberina
et al. Here we will expose and discuss the general calculation. \par

Radiative corrections for weak processes taking into account the unequal quark masses have
deserved much attention in the last years. In the nonleptonic three quark decay $b \to
q_1q_2\bar{q}_3$ with arbitrary quark masses, the calculation of the radiative corrections was
performed by Hokim and Pham \cite{11ref}. At lowest order, the current vertex had been computed
by Gavela et al. and Halprin et al. \cite{12r}, and by Paschalis and Gounaris and by Schilcher
et al. \cite{10bis}. The radiative correction to the semileptonic heavy quark decays was performed
by Cabibbo and Maiani and by Nir \cite{17r}. Recently, with the aim of applying the results to
Heavy Quark Effective Theory, the vertex has been reconsidered by Voloshin and Shifman
\cite{14ref} and by Neubert \cite{12bis}. \par

The paper is planned as follows. Although these are, as the folklore says,
``straightforward but tedious calculations'', we will give some details, helped with
Appendices in order to make the discussion simpler. In Section 2 we summarize the
calculation of the renormalized vertex and in Appendix III we compute the limits at
$q^2 = 0$ and at $q_{\max}^2$ to compare with the existing literature. In Section 3 we
give the corrected two-body rate. In Section 4 we summarize the calculation of the
Bremsstrahlung rate. In Section 5 we discuss the analytical results and the necessary
cancellations in ${1 \over D-4}$ and of the mass singularities, and we deduce the $m_q
\to 0$ and $q^2 \to 0$ limits of the final finite correction. In Section 6 we give
numerical results for several interesting cases, along with a discussion of the values of the quark masses.

\section{Renormalized vertex}
\hspace{\parindent} We use NDR regularization and follow the Feynman rules
and notations from the review article by Aoki et al. on the Standard Electroweak Model
\cite{11r}. In particular, Dirac algebra is performed in $D$ dimensions in Minkowski
space and we work in the Feynman gauge. \par

A long but straightforward calculation gives, for the renormalized
vertex
\bea
\label{20e}
&&\Lambda_{\mu}^{(R)} (p_q, p_b) = {16 \over 3} \ g_s^2 \ {1 \over 2^D \pi^{D/2}} \
\Gamma \left ( {6 - D \over 2} \right ) \left ( {m_b \over \mu} \right )^{D-4} \nn \\
&&\left [ \gamma_{\mu}\left ( 1 - \gamma_5 \right ) A_L(r, \xi ) + {p_{b \mu} \over m_b}
\left ( 1 + \gamma_5 \right ) B_+(r, \xi ) + {p_{q\mu} \over m_b} \left ( 1 + \gamma_5
\right ) C_+(r, \xi ) \right .\nn \\ 
&&\left . + \ \gamma_{\mu} \left ( 1 + \gamma_5 \right ) A_R(r, \xi ) + {p_{b \mu} \over
m_b} \left ( 1 - \gamma_5 \right ) B_-(r, \xi ) + {p_{q\mu} \over m_b} \left ( 1 -
\gamma_5 \right ) C_-(r, \xi ) \right ]
\eea

\noi where the coefficients of the different Dirac structures are~:
\bea
\label{21e}
&&A_L(r, \xi ) = {1 \over 2} \left [ - {D - 3 \over D - 4} \ I_1 + {(D-1)(1 + r^{D-4})
\over 2(D-4)(D-3)} - {1 - \xi + r^2 \over (D-4) (D-3)} \ I_2 \right .\nn \\
&&\left .  \qquad \qquad + {r^2 \over D-3} \ I_2 + {1 - r^2 \over D-3} \ I_3 \right ] \nn
\\ &&B_+(r, \xi ) = {1 \over 2} \ I_4 \nn \\
&&C_+ (r, \xi ) = - {1 \over 2} \left [ {5 - D \over D - 3} \ I_3 + I_4 \right ] \nn \\
&&A_R(r, \xi ) = {1 \over 2} \ r \ {1 \over D - 3} \ I_2 \nn \\
&&B_-(r, \xi ) = - {1 \over 2} \ r \left [ {2 \over D - 3} \ I_2 - {D-1 \over D-3} \ I_3
+ I_4 \right ] \nn \\
&&C_-(r, \xi ) = {1 \over 2} \ r \left ( I_2 - 2I_3 + I_4 \right ) \quad .
 \eea

\noi In the second term of the expression of $A_L(r, \xi )$ we recognize the vertex
counterterm, that we deduce from the Ward identity in Appendix I, and agrees with the result of
\cite{7r}. The quantities $I_i(\xi , r)$ are the integrals \bea
\label{22e}
&&I_1 = \int_0^1 dx \left [ r^2 (1 - x) + x - \xi x (1 - x) \right ]^{D/2-2} \nn \\
&&I_2 = \int_0^1 dx \left [ r^2 (1 - x) + x - \xi x (1 - x) \right ]^{D/2-3} \nn \\
&&I_3 = \int_0^1 dx \ x \left [ r^2 (1 - x) + x - \xi x (1 - x) \right ]^{D/2-3} \nn \\
&&I_4 = \int_0^1 dx \ x^2\left [ r^2 (1 - x) + x - \xi x (1 - x) \right ]^{D/2-3} 
    \eea

\noi whose expressions we give in Appendix II. \par

As a partial check of our calculation, we have computed the renormalized vertex at $q^2 = 0$
and at $q^2 = q^2_{max}$. We give our results in Appendix III. The renormalized vertex for
arbitrary masses and $q^2$ had also been computed by Paschalis and Gounaris \cite{10bis}, using
as infrared regulator a gluon mass $\lambda$, instead of dimensional regularization as we use
here. At $q^2 = 0$ it had been computed by Gavela et al. \cite{12r} and by Hokim and Pham
\cite{11ref}. For the IR finite form factors we agree with their result. At $q^2 = q^2_{max}$, we
agree with Paschalis and Gounaris \cite{10bis}, Voloshin and Shifman \cite{14ref}, and Neubert
\cite{12bis}. 

 \section{Two-body decay rate at order $\alpha_s$}
\hspace{\parindent} After a rather long calculation, one finds, for the two-body decay
rate at $O(\alpha_s)$~: 
\bea
\label{27e}
&&\Gamma_{D_s}^{Two-body} = \Gamma_{D_s}^{(0)} \left \{ 1 + {4 \over 3} \ {\alpha_s
\over \pi} \ g(D) \right . \nn \\
&&\left [ 2 A_L + {(1 - r^2)^2 - \xi^2 \over (1 - r^2)^2 - \xi (1 + r^2)}\ B_+ + {(1 -
\xi - r^2)^2 \over (1 - r^2)^2 - \xi (1 + r^2)} \ C_+ \right . \nn \\
&&+ {4r \xi \over (1 - r^2)^2 - \xi (1 + r^2) } \ A_R +
{r(1 + \xi - r^2)^2 \over ( 1 - r^2)^2 - \xi (1 + r^2)} \ B_- \nn \\
&&\left . \left .   + {r\left [ (1 - r^2)^2 -
\xi^2 \right ] \over (1 - r^2)^2 - \xi (1 + r^2)} \ C_- \right ] \right \}  \eea

\noi where $A_L, \cdots C_-$ are understood to be functions of $r$ and $\xi$, and the
function $g(D)$ is given by
\be
\label{28e}
g(D) = {\Gamma \left ( {6 - D \over 2} \right ) \over (2 \pi )^{D-4} (4 \pi)^{D-4}} \
{\Gamma \left ( {3 \over 2} \right ) \over \Gamma \left ( {D-1 \over 2} \right )}
\left ( {\pi m_b^2 \over \mu^2} \right )^{D-4} \left [ \lambda (1, r^2, \xi ) \right ]^{(D-4)/2}
\quad . \ee

\noi Expanding in powers of $D - 4$ one gets finally
\bea
\label{29e}
&&\Gamma^{Two-body}_{D_s} = \Gamma_{D_s}^{(0)} \left \{ 1 + {4 \over 3} \
{\alpha_s \over \pi} \ {1 \over (1 - r^2)^2 - \xi (1 + r^2)} \right . \nn \\ 
&&\left \{ 2\left ( A_L + D_L \right ) + {(1 - r^2)^2 - \xi^2 \over (1 - r^2)^2 - \xi (1
+ r^2)} \ B_+ + {(1 - \xi - r^2)^2 \over (1 - r^2)^2 - \xi (1 + r^2)} \ C_+ \right . \nn
\\
&&\left . + {4r \xi \over ( 1 - r^2)^2 - \xi (1 + r^2)} \ A_R + {r(1 + \xi - r^2)^2 \over
(1 - r^2)^2 - \xi (1 + r^2)} \ B_- + {r\left [ (1 - r^2)^2 - \xi^2 \right ] \over ( 1 -
r^2)^2 - \xi (1 + r^2)} \ C_- \right \} \nn \\ 
\eea

\noi where $D_L(r, \xi )$ comes from the expansion of $g(D)$ and the terms of order ${1
\over D - 4}$ in $A_L(r, \xi )$
\be
\label{30e}
D_L = \left [ 1 - {1 \over 2} (1 - \xi + r^2) I_2 \right ] \left \{ - 2 \log (2 \pi ) +
\gamma - 1 + \log \left ( {\pi m_b^2 \over \mu^2} \right ) + {1 \over 2} \log \left [ \lambda (1,
r^2, \xi ) \right ] \right \} \quad . \ee

\noi In this expression $I_2$ means the limit $I_2^{D=4}$ that can be read from the
formula in Appendix II. Analogously, the other terms become
\bea
\label{31e}
&&A_L \cong {1 \over 2} \Big \{ {1 \over D - 4} \left [ 3 - I_1 - (1 - \xi + r^2)I_2
\right ] - I_1 - 2 + {3 \over 2} \log (r) + (1 - r^2) I_3  \nn \\
&&\qquad  + (1 - \xi + 2r^2) I_2 \Big \} \nn \\
&&B_+ = {1 \over 2} \ I_4 \nn \\
&&C_+ = - {1 \over 2} \left ( I_3 + I_4 \right ) \nn \\
&&A_R = {1 \over 2} \ r \ I_2 \nn \\
&&B_- = - {1 \over 2} \ r \left ( 2 I_2 - 3I_3 + I_4 \right ) \nn \\
&&C_- = {1 \over 2} \ r \left ( I_2 - 2I_3 + I_4 \right )
 \eea

\noi where the limit $D \to 4$
 has been taken wherever it is possible, i.e. everywhere except in the first term of $A_L$.

\section{Bremsstrahlung rate}
\hspace{\parindent} From the expression of the real gluon emission amplitude
\bea
\label{32e}
&&{\cal M}_a^{Brem}(\lambda ) = {G \over \sqrt{2}} \ g_s \left [ \bar{u}\gamma_{\mu} (1
- \gamma_5) {{/ \hskip - 2 truemm p}_b - {/ \hskip - 2 truemm k} + m_b \over m_b^2 -
(p_b - k)^2} \ {/ \hskip - 2 truemm \varepsilon}^{(\lambda )} \ {\lambda_a \over 2} b
\ + \right . \nn \\ 
&&\left . \bar{u} \ {/ \hskip - 2 truemm \varepsilon}^{(\lambda )} \ {\lambda_a \over 2} \
{{/ \hskip - 2 truemm p}_u + {/ \hskip - 2 truemm k} + m_u \over m_u^2 - (p_u + k)^2} \
\gamma_{\mu} \ \left ( 1 - \gamma_5 \right ) b \right ] \bar{s} \gamma^{\mu} \left ( 1 -
\gamma_5 \right ) c
 \eea

\noi one gets the color and spin averaged rate for $D_s$ emission~:

\bea
\label{33e}
&&| \overline{{\cal M}^{Brem}}|^2 = - {1 \over 3} \ G^2 \ g_s^2 \ f_D^2 \ p_D^{\mu} \
p_D^{\nu} \nn \\
&&\left [ \bar{u} \gamma_{\mu} \left ( 1 - \gamma_5 \right ) {{/ \hskip - 2 truemm p}_b
- {/ \hskip - 2 truemm k} + m_b \over m_b^2 - (p_b - k)^2} \ \gamma_{\alpha} b + \bar{u}
\gamma_{\alpha} {{/ \hskip - 2 truemm p}_u + {/ \hskip - 2 truemm k} + m_u \over m_u^2 -
(p_u + k)^2} \ \gamma_{\mu} \left ( 1 - \gamma_5 \right ) b \right ] \nn \\
&&\left [ \bar{u} \gamma_{\nu} \left ( 1 - \gamma_5 \right ) {{/ \hskip - 2 truemm p}_b
- {/ \hskip - 2 truemm k} + m_b \over m_b^2 - (p_b - k)^2} \ \gamma^{\alpha} b + \bar{u}
\gamma^{\alpha} {{/ \hskip - 2 truemm p}_u + {/ \hskip - 2 truemm k} + m_u \over m_u^2 -
(p_u + k)^2} \gamma_{\nu} \left ( 1 - \gamma_5 \right ) b \right ]^+ \ .
 \eea

\noi Performing the Dirac algebra and putting quarks and gluons on-shell, one arrives at
the expression for the rate~: 
\bea
\label{34e}
&&\Gamma^{Brem} = - {2 \over 3} \ {1 \over (2 \pi )^{2D-3}} \ G^2 \ g_s^2 \ f_D^2 \ m_b^5
\nn \\
&&\left \{ \left [ (1 - r^2)^2 - \xi (1 + r^2)\right ] I^{-2,0} - {1 \over m_b^2}
\ 2 \left [ (1 - r^2)^2 - \xi (1 + r^2) \right ] I^{-1,0} \right . \nn \\
&&+ {1 \over m_b^4} \ 4(1 + r^2) I^{0,0} + {1 \over m_b^2} \ 2 \left [ (1 - r^2)^2
- \xi (1 + r^2) \right ] I^{0,-1} + r^2 \left [ (1 - r^2)^2 - \xi (1 + r^2) \right
] I^{0,-2} \nn \\
&&\left . - {1 \over m_b^4} \ 2(1 + r^2) I^{-1,1} - {1 \over m_b^4} \ 2(1 + r^2) I^{1,-1}
- (1 + r^2 - \xi) \left [ 1 - r^2)^2 - \xi (1 + r^2) \right ] I^{-1,-1} \right \} \nn \\
\eea

\noi where $I^{m,n}$ are the phase-space integrals in $D$ dimensions~:
\be
\label{35e}
I^{m,n} = \mu^{(m+n)(D-4)} \int {d^{D-1} p_u \over 2p_u^0} \ {d^{D-1} p_D \over 2 p_D^0} \
{d^{D-1} k \over 2k^0} \ \delta^D ( p_b - p_D - p_u - k ) \left (  p_b \cdot k \right
)^m \left ( p_u \cdot k \right )^n  \ee

\noi that are functions of $r$ and $\xi$ and whose expressions are given in Appendix IV.

\section{Discussion of the analytical results}

\subsection{$D_s$ rate}
\hspace{\parindent} It is convenient to reorganize the expressions of the two-body and
Bremsstrahlung decay rates in order to check the necessary cancellations, namely the ${1
\over D - 4}$ poles and the mass singularities. We can reorganize the expression for the
two-body rate in the form~:
\bea
\label{36e}
&&\Gamma^{Two-body}_{D_s} = \Gamma_{D_s}^{(0)} \left \{ 1 + {4 \over 3} \
{\alpha_s \over \pi} \left [ X + Y_L + {(1 - r^2)^2 - \xi^2 \over (1 - r^2)^2 -
\xi (1 + r^2)} \ B_+ \right . \right . \nn \\
&&+ {(1 - \xi - r^2)^2 \over (1 - r^2)^2 - \xi (1 + r^2)} \ C_+ + {4 \xi r \over (1 -
r^2)^2 - \xi (1 + r^2)} \ A_R \nn \\  
&&+ \left . \left .   {r (1 + \xi -
r^2)^2 \over (1 - r^2)^2 - \xi (1 + r^2)} \ B_- + {r \left [ (1 - r^2)^2 - \xi^2
\right ] \over (1 - r^2)^2 - \xi (1 + r^2)} \ C_- \right ] \right \}\eea

\noi where 
\bea
\label{37e}
&&X = \left [ 2 - {1 - \xi + r^2 \over \sqrt{\lambda (1, r^2, \xi )}} \log \left ( {1 +
r^2 - \xi + \sqrt{\lambda (1, r^2, \xi )} \over 1 + r^2 - \xi - \sqrt{\lambda (1, r^2,
\xi)}} \right ) \right ] \times \nn \\
&&\left [ {1 \over D-4} - 2 \log (2 \pi ) + \gamma +
\log \left ( {\pi m_b^2 \over \mu^2} \right ) \right ] \eea

\noi and the new quantity $Y_L$ is finite for $D \to 4$, given in Appendix V.
Analogously, we can rewrite the Bremsstrahlung rate in the form~:
\bea
\label{38e}
&&\Gamma_{D_s}^{Brem} = - \ \Gamma_{D_s}^{(0)} {4 \over 3} \ {\alpha_s \over \pi} \Big [
X + K^{-2,0} + r^2K^{0,-2} - \left ( 1 + r^2 - \xi \right ) K^{-1,-1} \nn \\ 
&&- 2K^{-1,0} + 2K^{0,-1} 
 + {4(1 + r^2) \over ( 1 - r^2)^2 - \xi (1 + r^2)} \ K^{0,0}\nn \\ 
&&- {2(1 + r^2) \over
(1 - r^2)^2 - \xi (1 + r^2)} \ K^{-1,1} - {2(1 + r^2) \over (1 - r^2)^2 - \xi (1 + r^2)}
K^{1,-1} \Big ] \eea

\noi where $X$ is the quantity defined by (\ref{37e}) and the expressions $K^{m,n}$ are
given in Appendix V. We observe that $X$, that contains the ${1 \over D-4}$ terms and the
terms in $\log \left ( {m_b \over \mu} \right )$, cancels between the two-body and the
Bremsstrahlung rates, as expected. To check the cancellation of the rest of the mass
singularities, we need to perform an expansion in powers of $m_q$ or in powers of $r$. From the
expression (\ref{36e}) and Appendix II we obtain, for $r \to 0$~:
\be
\label{39e}
\Gamma_{D_s}^{Two-body} = \Gamma_{D_s}^{(0)} \left \{ 1 + {4 \over 3} \ {\alpha_s \over
\pi} \left [ \Big [ D(r, \xi) \right ]_{r \to 0} + \left [ F_V(r, \xi ) \Big ]_{r
\to 0} \right ] \right \} \ee

\noi $[D(r, \xi )]_{r \to 0}$ contains the ${1 \over D-4}$ and the divergent terms as
$m_q \to 0$~:
\bea
\label{40e}
&&\left [ D(r, \xi ) \right ]_{r\to 0} = \left [ 2 - 2 \log \left ( {1 - \xi \over r}
\right ) \right ] \left [ {1 \over D-4} - 2 \log (2 \pi ) + \gamma + \log \left ( {\pi
m_b^2 \over \mu^2} \right ) \right ] \nn \\
&&+ \log (r) \left [ - {5 \over 2} + 2 \log (1 - \xi ) \right ] + \log^2(r)
\eea

\noi and $[F_V(r, \xi ) ]_{r \to 0}$ is the surviving finite piece~:
\bea
\label{41e}
&&\left [ F_V(r, \xi ) \right ]_{r \to 0} = - 3 - {\pi^2 \over 6} - 3 \log^2 (1 - \xi
) + 4 \log (1 - \xi ) + {1 \over \xi} \log (1 - \xi ) \nn \\
&&+ Sp(1 - \xi ) + \log \xi \log
(1 - \xi ) \quad .  \eea

\noi The Spence function $Sp(z)$ is defined in Appendix II. Analogously, we obtain~:
\be
\label{42e}
\Gamma_{D_s}^{Brem} = - \Gamma_{D_s}^{(0)} \ {4 \over 3} \ {\alpha_s \over \pi} \left
\{ [D(r, \xi )]_{r \to 0} + \left [ F_B(r, \xi ) \right ]_{r\to 0} \right \} \ee

\bea
\label{43e}
&&\left [ F_B(r, \xi ) \right ]_{r\to0} = - {21 \over 4} + {\pi^2 \over 3} + {13 \over 2} \log
(1 - \xi ) + \log (\xi ) \log ( 1 - \xi ) + Sp (\xi ) \nn \\
&&- 3 \log^2 (1 - \xi) + {\xi \over 1 - \xi} \log (\xi ) \quad . \eea

\noi We observe also that the singular terms in $\log (r)$ and $\log^2 (r)$ contained in
$[D(r, \xi )]_{r\to 0}$ cancel among the two-body and Bremsstrahlung rates. \par

Then, it follows, for $m_q \to 0$, the total rate~:
\bea
\label{44e}
&&\Gamma_{D_s}^{Two-body} + \Gamma_{D_s}^{Brem} = \Gamma_{D_s}^{(0)} \left \{ 1
+ {4 \over 3} \ {\alpha_s \over \pi} \left [ {9 \over 4} - {\pi^2 \over 3} -
2Sp(\xi ) - \log (\xi ) \log (1 - \xi )\right . \right .\nn \\
&&\left . \left . + {1 \over \xi} \log ( 1 - \xi ) - {5 \over 2} \log (1 - \xi ) - {\xi
\over 1 - \xi} \log (\xi ) \right ] \right \} \eea

\noi that gives, at $q^2 = 0$, i.e. $\xi = 0$~:
\be
\label{45e}
\Gamma_{D_s}^{Two-body} + \Gamma_{D_s}^{Brem} = \Gamma_{D_s}^{(0)} \left [ 1 + {4
\over 3} \ {\alpha_s \over \pi} \left ( {5 \over 4} - {\pi^2 \over 3} \right ) \right ]
\quad . \ee

\subsection{$D_s^*$ rate}
\hspace{\parindent} The calculation of the $D_s^*$ rate follows along the same lines, with the
replacement
\be
\label{46e}
f_D^2 \ p_D^{\mu} \ p_D^{\nu} \to f_{D^*}^2 m_{D^*}^2 \sum_{\lambda} \varepsilon^{(\lambda
)\mu} \ \varepsilon^{*(\lambda ) \nu} = f_{D^*}^2 p_{D^*}^{\mu} p_{D^*}^{\nu} -
f_{D^*}^2 m_{D^*}^2 g^{\nu \nu} \quad .  \ee

\noi The difference between the $D_s^*$ and $D_s$ rates is thus a term proportional to
$m_{D^*}^2$, i.e. $q^2$ or $\xi$. In the limit $\xi \to 0$ one must recover the same
decay rate. We obtain, for the two-body rate~:
\bea
\label{47e}
&&\Gamma_{D_s^*}^{Two-body} = \Gamma_{D_s^*}^{(0)} \left \{ 1 + {4 \over 3} \ {\alpha_s
\over \pi} \left [ X + Y_L + {\lambda (1, r^2, \xi ) \over (1- r^2)^2 + \xi (1 +r^2 -
2 \xi )}\ B_+ \right . \right . \nn \\
&&+ {\lambda (1, r^2, \xi ) \over (1 - r^2)^2 + \xi (1 + r^2 - 2 \xi )} \ C_+ - {12r \xi
\over (1 - r^2)^2 + \xi (1 + r^2 - 2 \xi )} \ A_R \nn \\
&&\left . \left . + {r \lambda (1, r^2, \xi ) \over (1 - r^2)^2 + \xi (1 + r^2 - 2 \xi)}\
B_- + {r\lambda (1, r^2, \xi ) \over (1 - r^2)^2 + \xi (1 + r^2 - 2 \xi)}\ C_- \right ]
\right \} \eea

\noi and for the Bremsstrahlung~:
\bea
\label{48e}
&&\Gamma_{D^*_s}^{Brem} = - \Gamma_{D^*_s}^{(0)} \ {4 \over 3} \ {\alpha_s \over \pi}
\left [ X + K^{-2,0} + r^2K^{0,-2} - (1 + r^2 - \xi ) K^{-1,-1} - 2K^{-1,0} 
+ \right . \nn \\
&&+ 2K^{0,-1} + {4 (1 + r^2) \over (1 - r^2)^2 + \xi (1 + r^2 - 2\xi )} \ K^{0,0} -
{2(1+r^2+2\xi ) \over (1 - r^2)^2 + \xi (1 + r^2 - 2\xi)} \ K^{-1,1} \nn \\
&&\left . - {2(1+r^2+2\xi
) \over (1 - r^2)^2 + \xi (1 + r^2 - 2\xi )} \ K^{1,-1} \right ] \quad . \eea

\noi Again, as expected, $X$, that contains the ${1 \over D-4}$ terms, cancels between
the two-body and the Bremsstrahlung rates. As for the mass singularities, we find in the case of
the $D_s^*$ exactly the same expression for $[D]_{r\to 0}$ that cancels among the two-body and
the Bremsstrahlung rates. The finite result is, in this case~:
\bea
\label{49e}
&&\Gamma_{D^*_s}^{Two-body} + \Gamma_{D^*_s}^{Brem} = \Gamma_{D^*_s}^{(0)} \left \{
1 + {4 \over 3} \ {\alpha_s \over \pi} \left [ 2 - {\pi^2 \over 3} - 2Sp (\xi ) - \log
(\xi ) \log (1 - \xi ) \right . \right . \nn \\
&&\left . \left . - {5 + 4 \xi \over 2(1 + 2 \xi )} \log (1 - \xi ) - {3 - \xi - 10 \xi^2
\over 4(1 - \xi ) (1 + 2\xi )} - {\xi (1 - \xi - 2\xi^2) \over (1 - \xi)^2 (1 + 2 \xi )}
\log (\xi ) \right ] \right \}  \eea

\noi that gives, at $\xi = 0$~:
\be
\label{50e}
\Gamma_{D^*_s}^{Two-body} + \Gamma_{D^*_s}^{Brem} = \Gamma_{D^*_s}^{(0)} \left [ 1 +
{4 \over 3} \ {\alpha_s \over \pi} \left ( {5 \over 4} - {\pi^2 \over 3} \right ) \right
] \ee

\noi i.e., the same correction than for the $D_s$, as expected.

 \section{Numerical results} 
\hspace{\parindent} To give the numerical results, let us parametrize the rate,
including the QCD corrections, under the form~:
\be
\label{51e}
\Gamma \left ( b \to D_s^-q \right ) = \Gamma^{(0)}\left  ( b \to D_s^-q \right ) \left [
1 + {4 \over 3} \ {\alpha_s \over \pi}\  \eta \left ( \xi_{D_s}, r \right ) \right ]
\ee

\noi and analogously for the $D_s^*$
\be
\label{52e}
\Gamma \left ( b \to D_s^{*-}q \right ) = \Gamma^{(0)}\left ( b \to D_s^{*-}q \right )
\left [ 1 + {4 \over 3} \ {\alpha_s \over \pi} \ \eta^* \left ( \xi_{D^*_s} , r 
\right ) \right ] \quad .\ee

\noi The functions $\eta(\xi, r)$ and $\eta^*(\xi , r)$ can be
read respectively from the finite sums of the two-body and Bremsstrahlung corrections,
formulae (\ref{36e}), (\ref{38e}) and (\ref{47e}), (\ref{48e}). The particular values
$\eta(\xi_{D_s} , r)$ and $\eta^*(\xi_{D^*_s}, r )$ correspond
res\-pec\-ti\-ve\-ly to both functions for $q^2 = m_{D_s}^2$ and $m_{D_s^*}^2$.  At fixed
$r$ these are slowly varying functions of $\xi$. For $r = 0$, these functions are given
by the relatively simple expressions (\ref{44e}) and (\ref{49e}), and for $r = r_c = {m_c
\over m_b}$ we use the full expressions. The functions {\it decrease} monotonously with
$\xi$, the dependence on $q^2$ is mild, and the values at the interesting values of low
$q^2$ (i.e. $\xi \simleq 0.2$) do not differ significantly from the $q^2 = 0$ value. For
$\xi \to \xi_{max}$ there is a logarithmic singularity, of the form $\log (1 - \xi )$ for
$r = 0$, smoothed out by the phase space $\Gamma^{(0)}(\xi_{max}) = 0$. The functions $\eta(\xi , 0)$, $\eta(\xi , r_c )$,
$\eta^*(\xi , 0 )$ and $\eta^*(\xi , r_c)$ decrease
by about 50 \% between $\xi = 0$ and $\xi = 0.6$ $\xi_{max}$. The correction depends
weakly on the ratio $r = {m_c \over m_b}$ in the expected range of quark
masses. The radiative corrections depend on the ratios $\xi_D$, $\xi_{D^*}$ and $r$ and
on $\alpha_s(\mu )$. We take $\mu = m_b$ and we adopt the value $\alpha_s(m_b) = 0.2$. \par

For the quark masses, we take pole masses from a fit to the semileptonic decay rate $b \to c
\ell^-\bar{\nu}_{\ell}$ with QCD corrections at one loop  \cite{17r}, to be consistent with the
same order that we compute here. The semileptonic decay rate reads, in this approximation~:
\be
\label{52ebis}
\Gamma (b \to c \ell^-\bar{\nu}_{\ell}) = {G^2 m_b^5 \over 192 \pi^3} \ |V_{cb}|^2 \  f_{PS}(r)
\left [ 1 + {4 \over 3} \ {\alpha_s \over \pi} \ f_{RC}(r) \right ] \ee

\noi where the phase space $f_{PS}(r)$ and radiative correction $f_{RC}(r)$ functions depend on $r
= {m_c \over m_b}$ and are given in \cite{17r}. Setting $r = 0.3$, we obtain, from the
semileptonic branching ratio 11 \%, \be
\label{52eter}
m_b = 4.85 \ \hbox{GeV} \qquad m_c = 1.45 \ \hbox{GeV} \quad .
\ee

\noi The mass difference $m_b - m_c = 3.40$~GeV compares well with the value $m_b - m_c = (3.43
\pm 0.04)$~GeV obtained in the $1/m$ expansion of the Heavy Quark Effective Theory \cite{2r}.
Therefore the pole masses that we choose (\ref{52eter}), at one loop, seem reasonable. With
these values, and $m_u \cong 0$, the results are the following, for the $u$-quark~:
\bea
\label{53e}
&&{4 \over 3} \ {\alpha_s \over \pi} \ \eta \left ( \xi_{D_S}, 0 \right ) = - 0.168 \nn
\\
&&{4 \over 3} \ {\alpha_s \over \pi} \  \eta^* \left ( \xi_{D_s^*}, 0 \right ) = - 0.159
   \eea

\noi and for the $c$-quark~:
\bea
\label{54e}
&&{4 \over 3} \ {\alpha_s \over \pi} \ \eta\left ( \xi_{D_S}, r_c \right ) = - 0.096 \nn
\\ &&{4 \over 3} \ {\alpha_s \over \pi} \ \eta^* \left ( \xi_{D_s^*}, r_c 
\right ) = - 0.108 \quad . \eea 

\noindent Then, using these values, the corrected branching ratios will be, from (\ref{8e}) and
(\ref{10e})~:
\bea
\label{55e}
&&BR \left ( b \to D_s^-c \right ) \cong 2.6 \ \% \nn \\
&&BR \left ( b \to D_s^{*-} c \right ) \cong 5.4 \ \% \quad .
\eea 

\noindent Our conclusion is that the sum including radiative corrections
\be
\label{56e}
BR \left ( b \to D_s^-c \right ) + BR \left ( b \to D_s^{*-} c \right ) \cong 8 \ \%
\ee

\noindent is still in agreement within $1\sigma$ with the measurement 
(\ref{11e}). Also, the
moderate radiative correction to the processes $b \to D_s^{(*)-}u$ reinforces our argument
about using the spectrum $\bar{B} \to D_s^-X_u$ in the measurement of the CKM matrix element
$V_{ub}$ \cite{1r}. \par

A number of remarks is in order here. First, from the relation at one loop between the pole mass
and the running $\overline{MS}$ mass

\be
m = \bar{m}(\bar{m}) \left [ 1 + {4 \over 3} \ {\alpha_s (\bar{m}) \over \pi} \right ] 
\label{47'}
\ee

\noi at first order in $\alpha_s$, from (\ref{52ebis}) one obtains~:

\be
\label{newformula1}
\Gamma (b \to c \ell \bar{\nu}_{\ell}) = {G^2[\bar{m}_b(\bar{m}_b)]^5 \over 192 \pi^3} \
|V_{cb}|^2 \ f_{PS}(r) \left \{ 1 + {4 \over 3} \ {\alpha_s (\bar{m}_b) \over \pi} \ \left [ 5 +
f_{RC}(r) \right ]\right \} \ee

\noi and in formulas (\ref{51e}) and (\ref{52e}) when $[\bar{m}_b(\bar{m}_b)]^3$ is
substituted to $m_b^3$, a term ${4 \over 3} {\alpha_s \over \pi} \times 3$ has to be
added. Taking $\alpha_s(\bar{m}_b) = 0.22$ we get from (\ref{47'})

\be
\label{newformula2}
\bar{m}_b (\bar{m}_b) = 4.43 \ \hbox{GeV} \quad .
\ee

\noi The values of the pole mass (\ref{52eter}) and of the $\overline{MS}$ running mass
(\ref{newformula2}) are respectively smaller and larger than the values recently quoted in the
literature from the analysis of semileptonic $b$ decay, the reason being that a partial
resummation of higher oder diagrams is made that enhances the radiative corrections by roughly
a factor 2 (see for example \cite{2r}). The interest of considering the $\overline{MS}$ mass is
that the series is Borel summable, while using the pole mass, there is a renormalon
ambiguity, cancelled by another renormalon ambiguity in the pole mass.
Ball et al. \cite{2r} quote, as central values, $m_b = 5.05$~GeV, $m_c = 1.62$~GeV,
and $\bar{m}_b(\bar{m}_b) = 4.23$~GeV, $\bar{m}_c(\bar{m}_c) = 1.29$~GeV, leading to
consistent results for the semileptonic rate in both the $\overline{MS}$ and on-shell
schemes. \par

Concerning the radiative corrections at higher orders in the processes $b \to D_s^{(*)-}c$,
one part of the radiative corrections has the same topology than in semileptonic decays
(simply at a given value of $q^2$ instead of integrating over $q^2$), and we could expect that
these radiative corrections would be similarly enhanced by higher orders. This is by the way
what happens at order $\alpha_s$~: the correction that we obtain for $b \to D_s^{(*)-} c$ is
very close to the one obtained at the same order in the semileptonic case \cite{17r}. This would
imply, e.g. in the on-shell renormalization scheme, a larger radiative correction by a factor 2
\cite{2r}, but consistently also a larger $m_b = 5.05$~GeV, leading grosso modo to the same
results (\ref{55e}) and (\ref{56e}). We must keep in mind however that, already at second order in
$\alpha_s$, we could have another type of corrections, absent in the semileptonic decay, that
break factorization (e.g. two or more gluons linking the $D_s$ to the quark legs). We can only hope
that these corrections will be small, like it is the case for the short distance QCD
coefficient (\ref{4e}), that empirically is very close to 1. \par 

Work remains to be done, in particular the calculation of the QCD-corrected spectrum $b \to
D_s^-c$ to be compared with the measured spectrum \cite{4ref} of $B \to D_s^{\pm}X$. It would be
interesting to check in particular if the same function describing the $b$-quark Fermi motion
\cite{14r} fits the semileptonic spectrum and the inclusive $D_s$ one as well.

\section*{Acknowledgements}
\hspace{\parindent} The authors acknowledge useful discussions with J. Charles, and partial
support from the EEC-TMR Program, contract N. CT98-0169.

\newpage
\section*{Appendix I. On-shell renormalization, Ward identity and vertex counterterm}
\hspace{\parindent} The bare self-energy of a quark of mass $m$ writes
$$\Sigma (p) = A(p^2) {/ \hskip - 2 truemm p} + B (p^2)$$

\noi with
\begin{eqnarray*}
&&A(p^2) = {4 \over 3} \ g_s^2 \ {1 \over 2^D \pi^{D/2}} \ \Gamma \left ( {4 - D \over
2} \right ) m^{D-4} \ {4 \over D} \ F\left ( {4 - D \over 2}, 2 ; {D \over 2} + 1 ; \xi
\right )  \\  
&&B(p^2) = - {4 \over 3} \ g_s^2 \ {1 \over 2^D \pi^{D/2}} \ \Gamma \left ( {4 - D \over 
2} \right ) m^{D-4} \ {2D \over D - 2} \ m\ F\left ( {4 - D \over 2}, 1 ; {D \over 2} 
; \xi \right ) \ . 
\end{eqnarray*}

\noi To proceed with the on-shell renormalization it is convenient to expand $\Sigma
(p)$ in powers of ${/ \hskip - 2 truemm p} - m$, that gives~: 
$$\Sigma (p) = - {4 \over 3} \ g_s^2 \ {1 \over 2^D \pi^{D/2}} \ \Gamma \left ( {4 - D
\over 2} \right ) \ m^{D-4} \left [ {2(D+1) \over D - 3} \ m + {1 - D \over D - 3} \ {/
\hskip - 2 truemm p} \right ] + \cdots$$

\noi The unknowns $a$, $b$ in the self-mass counterterm
$$\sigma (p) = a {/ \hskip - 2 truemm p} + b$$

\noi will be fixed by the on-shell renormalization conditions \cite{11r}~:
$$\left . \bar{u}(p) \ \Sigma^{(R)} (p) \right |_{{/ \hskip - 1 truemm p} = m} = 0 \qquad
\left . \Sigma^{(R)}(p) \ u(p) \right |_{{/ \hskip - 2 truemm p} = m} = 0$$

\noi where $u(p)$ is a solution of the Dirac equation with mass $m$~: $({/ \hskip - 2
truemm p} - m) \ u(p) = 0$, and $\Sigma^{(R)} (p) = \Sigma (p) + \sigma (p)$. These
conditions yield
$$\left . \sigma (p) = {4 \over 3} \ g_s^2 \ {1 \over 2^D \pi^{D/2}} \ \Gamma \left ( { 4 -
D \over 2} \right ) m^{D-4} \left [ {2(D+1) \over D-3} \ m + {1 - D \over D - 3} \ {/
\hskip - 2 truemm p} \right ] + \cdots \right ) \quad .$$

\noi On the other hand, as we use NDR, that preserves chiral symmetry, both the
self-energy $\sigma (p)$ and current $\lambda_{\mu}^{qb}$ counterterms will satisfy
separately the Ward identity (an overall factor ${g \over 2 \sqrt{2}} V_{qb}$ is
understood, where $g$ is the weak coupling) \cite{11r} \cite{11bis}~:
$$\left ( p_q - p_b \right )^{\mu} \ \lambda_{\mu}^{qb} = \left [ \sigma^q (p_q) T^{(+)}
\left ( 1 - \gamma_5 \right ) - \left ( 1 + \gamma_5 \right ) T^{(+)} \ \sigma^b (p_b)
\right ] - i M_W \ \lambda^{qb}$$

\noi where $\lambda^{qb}$ is the counterterm of the coupling of the unphysical Higgs and
$T^{(+)}b = q$. One gets immediately
$$\lambda_{\mu}^{ub} = {4 \over 3} \ g_s^2 \ {1 \over 2^D \pi^{D/2}} \ \Gamma \left ( {6 -
D \over 2} \right ) {D - 1 \over (D - 3)(D - 4)} \left [ m_q^{D-4} + m_b^{D-4} \right ]
\gamma_{\mu} \left (1 - \gamma_5 \right ) \quad .$$

\vskip 1 truecm 
\section*{Appendix II. Vertex integrals}
\hspace{\parindent} An expansion is made of the vertex integrals $I_i(\xi , r)$ up to first power
of $D-4$ in the cases in which this is necessary ($I_1$ and $I_2$)~:
 \begin{eqnarray*}
&&I_1 = 1 - {D-4 \over 2} \left [ 2 + {1 - r^2 - \xi \over 2 \xi } \log (r^2) +
{\sqrt{\lambda (1,r^2, \xi )} \over 2 \xi} \log \left ( {1 + r^2 - \xi + \sqrt{\lambda
(1, r^2, \xi )} \over 1 + r^2 - \xi - \sqrt{\lambda (1, r^2, \xi )}} \right ) \right ] \\
&& \\
&&I_2 = {1 \over \sqrt{\lambda (1, r^2, \xi)}} \log \left ( {1 + r^2 - \xi +
\sqrt{\lambda (1, r^2, \xi )} \over 1 + r^2 - \xi - \sqrt{\lambda (1, r^2, \xi )}}
\right ) \left [ 1 + {1 \over 2} (D-4) \log \xi \right ] \\
&&+ {1 \over 2} (D - 4) {1 \over \sqrt{\lambda (1, r^2, \xi)}} \left [ - 2 Sp \left ( {2
\sqrt{\lambda (1,r^2, \xi)} \over 1 - r^2 + \xi + \sqrt{\lambda (1, r^2, \xi )}} \right
) + 2 Sp \left ( {2 \sqrt{\lambda (1,r^2, \xi )} \over 1 - r^2 - \xi + \sqrt{\lambda (1,
r^2, \xi)}} \right ) \right . \\
&&- {1 \over 2} \log^2 \left ( {1 - r^2 + \xi - \sqrt{\lambda (1, r^2, \xi )} \over 1 -
r^2 + \xi + \sqrt{\lambda (1, r^2,\xi )}} \right ) + {1 \over 2} \log^2 \left ( {1 -r^2 -
\xi - \sqrt{\lambda (1, r^2, \xi )} \over 1 - r^2 - \xi + \sqrt{\lambda (1, r^2, \xi )}}
\right ) \\
&&+ \log^2 \left ( {1 - r^2 + \xi - \sqrt{\lambda (1, r^2, \xi )} \over 2 \xi} \right ) -
\log^2 \left ( {1 - r^2 - \xi - \sqrt{\lambda (1,r^2,\xi)} \over 2 \xi}\right ) \\
&&\left . - \log^2 \left ( {1 - r^2 + \xi + \sqrt{\lambda (1, r^2, \xi )} \over 2 \xi}
\right ) + \log^2 \left ( {1 - r^2 - \xi + \sqrt{\lambda (1,r^2,\xi)} \over 2 \xi}\right
) \right ]\\
&& \\
&&I_3 = - {1 \over 2 \xi} \log (r^2) - {1 - r^2 - \xi \over 2 \xi \sqrt{\lambda (1,
r^2, \xi )}} \log \left ( {1 + r^2 - \xi + \sqrt{\lambda (1, r^2, \xi )} \over 1 + r^2
- \xi - \sqrt{\lambda (1, r^2, \xi )}} \right ) \\
&& \\ 
&&I_4 = {1 \over \xi} + {1 - r^2 - \xi \over 2 \xi^2} \log (r^2) + {(1 - r^2)^2 + \xi
(\xi - 2) \over 2 \xi^2 \sqrt{\lambda (1,r^2, \xi )}} \log \left ( {1 + r^2 - \xi +
\sqrt{\lambda (1, r^2, \xi )} \over 1 + r^2 - \xi - \sqrt{\lambda (1, r^2, \xi)}} \right
) \quad .
\end{eqnarray*}

\vskip 3 truemm

The Spence function $Sp(z)$ is defined by
\begin{eqnarray*}
&&Sp(z) = \int_0^1 dt {\log (t) \over t - {1 \over z}} \\
&&Sp(z) = \sum_{n=1}^{\infty} {z^n \over n^2} \qquad , \quad |z| \leq 1 \quad . 
\end{eqnarray*}

\noi In the limit $q^2 \to q_{max}^2$, since $\xi_{max} \to (1 - r)^2$ and $\lambda (1,
r^2, \xi_{max}) \to 0$, one obtains~:
\begin{eqnarray*}
&&I_1(\xi_{max}, r) = 1 - (D-4) \left [ 1 + {r \over 1 - r} \log (r) \right ] \\
&&I_2(\xi_{max}, r) = {1 \over r} \left [ 1 + (D-4) \left ( 1 + {1 \over 1 - r} \log (r)
\right )\right ] \\ 
&&I_3(\xi_{max}, r) = - {1 \over 1 - r} \left [{1 \over 1 - r} \log (r) + 1 \right ] \\
&&I_4(\xi_{max}, r) = {1 + r \over (1 - r)^2} + {2r \over (1 - r)^3} \log (r) \quad .
\end{eqnarray*}

\noindent Another limit used in the text is $q^2$ or $\xi \to 0$~:
\begin{eqnarray*}
&&I_1(0, r) = 1 - {D-4 \over 2} \left [ 1 + {1 \over 1 - r^2} \log (r^2) - \log (r^2) \right ]
\\ 
&&I_2(0, r) = - {1 \over 1 - r^2} \log (r^2) - {1 \over 4} (D-4) {1 \over 1 - r^2}
\log^2 (r^2)  \\ 
&&I_3(0, r) =  {1 \over 1 - r^2} + {r^2 \over (1 - r^2)^2} \log (r^2)  \\
&&I_4(0, r) = - {r^4 \over (1 - r^2)^3} \log (r^2) + {1 - 3r^2 \over 2(1 - r^2)^2} 
\quad . \end{eqnarray*}

\vskip 1 truecm
\section*{Appendix III. Renormalized vertex at $q^2 = 0$ and $q^2 = q^2_{max}$}
\subsection*{Vertex at $q^2 = 0$}
\hspace{\parindent} Defining the quark couplings by the expansion of the renormalized vertex $(q = p_b - p_u = p - p')$~: 
$$\Lambda_{\mu}^{(R)}(p', p) = g_V \gamma_{\mu} - g_A \gamma_{\mu} \gamma_5 + {g_M \over
2m_b} \ i\sigma_{\mu \nu}q^{\nu} + {g_T\over 2 m_b} \ i \sigma_{\mu \nu} q^{\nu}\gamma_5 
+ {g_S  \over 2m_b} \ q_{\mu} + {g_P \over 2 m_b} \ q_{\mu}  \ \gamma_5$$

\noindent we find, at $q^2 = 0$~: 
\begin{eqnarray*}
&&g_V = 1 + {4 \over 3} \ {\alpha_s \over \pi} \left \{ - {5 \over 8} + {1 \over 16}\  {1 + r^2
\over 1 - r^2} \log^2 (r^2) + {- 3 - 7r^2 - 8r + 2(1 + r)^2 \over 16(1 - r^2)} \log (r^2) \right .
 \\
&&\left . + {1 \over 2} \left [ {1 \over D - 4} + \log \left ( {\pi m_b^2 \over \mu^2} \right ) +
\gamma - 2 \log (2 \pi ) \right ] \left [ 2 + {1 + r^2 \over 1 - r^2} \log (r^2) \right ] \right
\}  \\
&& \\
&&g_A  = 1 + {4 \over 3} \ {\alpha_s \over \pi} \left \{ - {5 \over 8} + {1 \over 16} \ {1 + r^2
\over 1 - r^2} \log^2 (r^2) + {- 3 - 7r^2 + 8r + 2(1 - r)^2 \over 16(1 - r^2)} \log (r^2) \right .
 \\ 
&&\left . + {1 \over 2} \left [ {1 \over D - 4} + \log \left ( {\pi m_b^2 \over \mu^2} \right ) +
\gamma - 2 \log (2 \pi ) \right ] \left [ 2 + {1 + r^2 \over 1 - r^2} \log (r^2) \right ] \right
\}  \\
&&  \\
&&g_M = {4 \over 3} \ {\alpha_s \over \pi} \left [ - {1 - r \over 2(1 - r^2)} + {r(1 - r) \over
2(1 - r^2)^2} \log (r^2) \right ]  \\
&&  \\
&&g_T = {4 \over 3} \ {\alpha_s \over \pi} \left [ - {1 + r \over 2(1 - r^2)} - {r(1 + r) \over
2(1 - r^2)^2} \log (r^2) \right ]  \\
&& \\
&&g_S = {4 \over 3} \ {\alpha_s \over \pi} \left [  {1  \over (1 - r^2)^2} (1 + 2r - 2r^2 - r^3) +
{r \over 2(1 - r^2)^3} (3 + r - r^2 - 3r^3) \log (r^2) \right ]  \\
&& \\
&&g_P = {4 \over 3}\  {\alpha_s \over \pi} \left [  {1  \over (1 - r^2)^2} (1 - 2r - 2r^2 + r^3) -
{r \over 2(1 - r^2)^3} (3 - r - r^2 + 3r^3) \log (r^2) \right ] \ .  \\ 
\end{eqnarray*}

\noindent The finite couplings $g_M$, $g_T$, $g_S$, $g_P$ agree with former calculations by
Halprin et al., and by Gavela et al. \cite{12r}, and by Hokim and Pham \cite{11ref}. We find
infrared divergent results for $g_V$, $g_A$, as we found in \cite{12r} with a gluon mass
$\lambda$ as infrared cut-off. In our expressions for $g_V$ and $g_A$ the last term, that
vanishes for $r \to 1$, cancels when the two-body decay rate is added to the Bremsstrahlung rate,
just as it happened with the $\lambda$ regulator. In the equal mass limit $r \to 1$, we find,
keeping fixed the infrared cut-off, the expected result~:
$$\Lambda_{\mu}^{(R)}(p',p) = \gamma_{\mu} \left ( 1 - \gamma_5 \right ) + {4 \over 3}
\ {\alpha_s \over \pi} \left [ {1 \over 2} \gamma_{\mu} \ \gamma_5 - {1 \over 2} \ {1
\over 2m} \ i \sigma_{\mu \nu} \left ( p^{\nu} - p'^{\nu} \right ) + {7 \over 6} \ {1
\over 2m} \left ( p_{\mu} - p'_{\mu} \right ) \gamma_5 \right ] \ .$$

\vskip 5 truemm
\subsection*{Vertex at $q^2 = q_{max}^2$}
\hspace{\parindent} From (\ref{21e}) and the limit of the integrals (\ref{22e}) at $q^2 \to
q_{max}^2$ given in Appendix II, we obtain 
\begin{eqnarray*}
&&\Lambda_{\mu}^{(R)}(p',p) = \gamma_{\mu} \left \{ 1 + {4 \over 3} \ {\alpha_s \over \pi} \left
[ - 1 - {1 + r \over 2(1 - r)} \log (r) \right ] \right \} \\ 
&&- \gamma_{\mu} \gamma_5 \left \{ 1 + {4 \over 3} \ {\alpha_s \over \pi} \left [ - {3 \over 2} - {1 + r \over 2(1 - r)} \log (r)
\right ] \right \} \\ 
&&+ {4 \over 3} \ {\alpha_s \over \pi} \left \{ {1 \over 2m_b} \ i \sigma_{\mu \nu}
q^{\nu} \left [ {1 \over 2(1 -r)} \log (r) \right ] + {1 \over 2m_b} \ i \sigma_{\mu \nu}
q^{\nu} \gamma_5 \left [ {1 \over 1 -r } + {1 + r \over 2(1 - r)^2} \log (r) \right ] \right . \\
&&\left . + {q_{\mu} \over 2m_b} \left [ - {1 \over 1 - r} - {1 + r \over 2(1 - r)^2} \log (r)
\right ] + {q_{\mu} \over 2 m_b} \ \gamma_5 \left [ {2(1 + r) \over (1 - r)^2} - {1 + r^2 - 10 r
\over 2(1 - r)^3} \log (r) \right ] \right \} \end{eqnarray*}

\noi in agreement with Paschalis and Gounaris \cite{10bis}. In terms of four-velocities, this
reads, \begin{eqnarray*}
&&\Lambda_{\mu}^{(R)}(p,p') = \gamma_{\mu} \left \{ 1 + {4 \over 3} \ {\alpha_s \over \pi} \left
[ - 1 -{3 \over 4} \  {1 + r \over 1 - r} \log (r) \right ] \right \} \\
&&- \gamma_{\mu} \gamma_5 \left \{ 1 + {4 \over 3} \ {\alpha_s \over \pi}\left [ - 2 - {3 \over 4}  \ {1 + r \over 1 - r} \log (r)
\right ] \right \} \\ 
&&+ {4 \over 3} \ {\alpha_s \over \pi} \left \{ v_{\mu} \left [ - {1 \over 2(1 - r)} - {r \over
2(1 - r)^2} \log (r) \right ] + v'_{\mu} \left [ {r \over 2(1 - r)} + {r \over 2(1 - r)^2} \log
(r) \right ] \right . \\
&&\left . + v_{\mu} \gamma_5 \left [  {3 + r \over 2(1 - r)^2} + {r(5 -  r) \over 2(1 - r)^3}
\log (r) \right ] + v'_{\mu} \gamma_5 \left [ - {r(1+ 3r \over 2(1 - r)^2} - {r(5r -1)
\over 2(1 - r)^3} \log (r) \right ] \right \} 
 \end{eqnarray*}

\noi that is also in agreement with the calculation of Neubert (first reference \cite{12bis}).
Finally, using the Gordon identities for unequal masses \begin{eqnarray*}
&&\bar{u}(p') \gamma_{\mu} \ u(p) = {1 \over 2} \ \bar{u}(p') \left [ v_{\mu} + v'_{\mu} -
i \sigma_{\mu \nu} \left ( v^{\nu} - v'^{\nu} \right ) \right ] u(p) \\ 
&&\bar{u}(p') \left [ v_{\mu} + v'_{\mu} - i \sigma_{\mu \nu} \left ( v^{\nu} -
v'^{\nu} \right ) \right ] \gamma_5 \ u(p) = 0\end{eqnarray*}

\noi one obtains the form
\begin{eqnarray*}
&&\Lambda_{\mu}^{(R)}(p',p) = \gamma_{\mu} \left ( 1 - \gamma_5 \right ) \nn \\
&&+ {4 \over 3} \ {\alpha_s \over \pi} \left \{ \gamma_{\mu} \left [ - {3 \over 2}
- {3 \over 4} \ {1 + r \over 1 - r} \log (r) \right ] - \gamma_{\mu} \gamma_5
\left [ - 2 - {3 \over 4} \ {1 + r \over 1 - r} \log (r) \right ] \right .  \\
&&- {1 \over 2} \ {1 \over 2} i \sigma_{\mu \nu}\left ( v^{\nu} - v'^{\nu} \right ) + {1
\over 2} \ i \sigma_{\mu \nu} \left ( v^{\nu} - v'^{\nu} \right ) \gamma_5 \left [ {3 (1
+ r) \over 2(1 - r)} + {3r \over (1 - r)^2} \log (r) \right ] \\
&&+ {1 \over 2} \left ( v_{\mu} - v'_{\mu} \right ) \left [ - {1 + r \over 2(1 - r)} - {2r
\over 2(1 - r)^2} \log (r) \right ]  \\
&&\left . + {1 \over 2} \left ( v_{\mu} - v'_{\mu} \right ) \gamma_5 \left [ {3r^2 + 2r +
3 \over 2(1 - r)^2} + {4r (1 + r) \over 2 (1 - r)^3} \log (r) \right ] \right \} \quad . 
\end{eqnarray*}

\noi The vector and axial vector couplings have been written down by Neubert \cite{12bis}, and
we agree with his result. In the equal mass limit $r \to 1$ we recover, as we must, the
result of equal masses at $q^2 = 0$ given above.

\vskip 1 truecm
\section*{Appendix IV. Phase space integrals}
\hspace{\parindent} Two-body phase space
\begin{eqnarray*}
&&I_0 = \mu^{4-D}\int {d^{D-1} p_u \over 2p_u^0} \ {d^{D-1} p_D \over 2 p_D^0} \ \delta^D
 ( p_b - p_D - p_u )  \\
&& \\  
&&I_0 = {\pi^{(D-1)/2} \over 2^{D-2}} \ {1 \over \Gamma \left ( {D-1 \over 2}\right ) }
\left ( { m_b \over \mu} \right )^{D-4} \left [ \lambda (1, r^2, \xi ) \right ]^{(D-3)/2} \quad .
 \end{eqnarray*}

\noi Three-body phase space integrals
\begin{eqnarray*}
&&I^{m,n} = \mu^{(m+n)(D-4)} \int {d^{D-1} p_u \over 2 p_u^0} \ {d^{D-1} p_D \over 2p_D^0} \
{d^{D-1} k \over 2k^0} \ \delta^D ( p_b - p_D - p_u - k ) \left ( p_b \cdot k
\right )^m \left ( p_u \cdot k \right )^n \\
&& \\
&& \\
&&I^{-2,0} = \pi^2 {1 \over m_b^2} \left \{ {1 \over D-4} \sqrt{\lambda (1, r^2, \xi )}
\right . \\
 &&+ \left [ \log \left ( {\pi m_b^2 \over \mu^2} \right ) + \gamma - 2 + 3 \log \left (
\sqrt{\lambda (1, r^2, \xi )} \right ) - \log (r) - \log \left ( \xi^{1/2} \right )
\right ] \sqrt{\lambda (1,r^2,\xi )} \\
&&\left . - \left ( 1 - \xi + r^2 \right ) \log \left ( {1 + r^2 - \xi + \sqrt{\lambda
(1, r^2, \xi )} \over 2r} \right ) - \left ( 1 + \xi - r^2 \right ) \log \left ( {1 -
r^2 + \xi + \sqrt{\lambda (1, r^2, \xi )} \over 2 \xi^{1/2}} \right ) \right \} \\
&& \\
&& \\
&&I^{0,-2} = \pi^2 {1 \over m_b^2} \ {1 \over r^2} \left \{ {1 \over D-4} \sqrt{ \lambda
(1, r^2, \xi)} \right .\\
&&+ \left [ \log \left ( {\pi m_b^2 \over \mu^2} \right ) + \gamma - 2 + 3 \log \left (
\sqrt{\lambda (1, r^2, \xi )} \right ) - \log (r) - \log \left ( \xi^{1/2} \right ) \right ]
\sqrt{\lambda (1, r^2, \xi )} \\
&&-\left .  2(1 - \xi ) \log \left ( {1 + r^2 - \xi + \sqrt{\lambda (1, r^2, \xi )} \over
2r} \right ) - \left ( 1 - \xi -r^2 \right ) \log \left ( {1 - r^2 + \xi + \sqrt{\lambda
(1, r^2, \xi )} \over 2 \xi^{1/2}} \right ) \right \} \\
&& \\
&& \\
&&I^{-1,-1} = \pi^2 {1 \over m_b^2} \left \{ {2 \over D-4} \log \left ( {1 + r^2 - \xi +
\sqrt{\lambda (1, r^2, \xi)} \over 2r} \right ) \right . \\
&&+ \left [ 2 \log \left ( {\pi m_b^2 \over \mu^2} \right ) + 2 \gamma - 2 + 4 \log (r) + \log
\left ( {1 - r^2 + \xi + \sqrt{\lambda (1, r^2, \xi )} \over 2 \xi} \right ) \right .\\
&&+ \left . 3 \log \left ( {1 + r^2 - \xi + \sqrt{\lambda (1, r^2, \xi )} \over 2r} \right
) \right ] \log \left ( {1 + r^2 - \xi + \sqrt{\lambda (1, r^2, \xi)} \over 2r} \right )
\\
&&+ \log (r) \log \left ( {1 - r^2 + \xi + \sqrt{\lambda (1, r^2, \xi )} \over 2 \xi }
\right ) - {\pi^2 \over 3} + 3 Sp \left  [ \left ( {1 + r^2 - \xi - \sqrt{\lambda (1,
r^2, \xi )} \over 2r} \right )^2 \right ] \\
&&\left . - Sp \left [ {1 - r^2 + \xi - \sqrt{\lambda (1, r^2, \xi )} \over 2} \right
] - Sp \left [ {1 + r^2 - \xi - \sqrt{\lambda (1, r^2, \xi)} \over 2} \right ] 
\right \} \\
&& \\
&& \\
&&I^{-1,0} = \pi^2 \left \{ {1 \over 2} \sqrt{\lambda (1, r^2, \xi )} - r^2 \log \left
( {1 + r^2 - \xi + \sqrt{\lambda (1, r^2, \xi )} \over 2r} \right ) \right . \\
&&\left . - \xi \log \left ( {1 - r^2 + \xi + \sqrt{\lambda (1, r^2, \xi )} \over 2
\xi^{1/2}} \right ) \right \} \\
&& \\
&& \\
&&I^{0,-1} = \pi^2 \left \{ - {1 \over 2} \sqrt{\lambda (1, r^2, \xi )} + (1 - \xi )
\log \left ( {1 + r^2 - \xi + \sqrt{\lambda (1, r^2, \xi)} \over 2r} \right ) \right .\\
&&\left . - \xi \log \left ( {1 - r^2 + \xi + \sqrt{\lambda (1, r^2, \xi )} \over 2
\xi^{1/2}} \right ) \right \} \\
&& \\
&& \\
&&I^{-1,1} = \pi^2 \ m_b^2 \ {1 \over 4}\left \{ {1 \over 4} \left ( 1 - 3r^2 + 5 \xi
\right ) \sqrt{\lambda (1, r^2, \xi )} \right . \\
&&+ \left . r^4 \log \left ( {1 + r^2 - \xi + \sqrt{\lambda (1, r^2, \xi)} \over 2r}
\right ) - \xi \left ( 2 - 2r^2 + \xi \right ) \log \left ( {1 - r^2 + \xi +
\sqrt{\lambda (1, r^2, \xi )} \over 2 \xi^{1/2}} \right ) \right \} \\
&& \\
&& \\
&&I^{1,-1} = \pi^2 \ m_b^2 \ {1 \over 8} \left \{ - {1 \over 2} \left ( 3 - r^2 - 5 \xi
\right ) \sqrt{\lambda (1, r^2, \xi )} \right . \\
&&+ 2 \left [ ( 1 - \xi)^2 + 2r^2 \xi \right ] \log \left ( {1 + r^2 - \xi +
\sqrt{\lambda (1, r^2, \xi )} \over 2r} \right ) \\
&&\left . - 2 \xi \left ( 2 - 2r^2 - \xi \right ) \log \left ( {1 - r^2 + \xi +
\sqrt{\lambda (1, r^2, \xi)} \over 2 \xi^{1/2}} \right ) \right \} \\ &&
\\
&&I^{0,0} = \pi^2 \ m_b^2 \ {1 \over 4} \ \left \{ {1 \over 2} \left ( 1 + r^2 + \xi
\right ) \sqrt{\lambda (1, r^2, \xi )} \right .\\
&&\left . - 2r^2 (1 - \xi ) \log \left ( {1 + r^2 - \xi + \sqrt{\lambda (1, r^2, \xi )}
\over 2r} \right ) - 2 \xi (1 - r^2) \log \left ( {1 - r^2 + \xi + \sqrt{\lambda (1, r^2,
\xi )} \over 2 \xi^{1/2}} \right ) \right \} \  . \end{eqnarray*}
 
\vskip 1 truecm
\section*{Appendix V}
\hspace{\parindent} Here we define necessary expressions that enter into formulae
(\ref{36e}), (\ref{38e}) and (\ref{47e}), (\ref{48e}), that are not given in the course
of the text. For the two-body decay rate we have the new expression~: 
\begin{eqnarray*}
&&Y_L = \left [ 2 - {1 - \xi + r^2 \over \sqrt{\lambda (1, r^2, \xi )}} \log \left ( {1
+ r^2 - \xi + \sqrt{\lambda (1, r^2, \xi )} \over 1 + r^2 - \xi - \sqrt{\lambda (1, r^2,
\xi)}} \right ) \right ] \left \{ - 1 + {1 \over 2} \log \left [ \lambda (1, r^2, \xi )
\right ] \right \} \\
&&- 2 + {1 - r^2 - \xi \over 4 \xi} \ \log (r^2) + {\sqrt{\lambda (1, r^2, \xi)} \over
4 \xi} \ \log \left ( { 1 + r^2 - \xi + \sqrt{\lambda (1, r^2, \xi)} \over 1 + r^2 -
\xi - \sqrt{\lambda (1, r^2, \xi )}} \right ) \\
&&+ {3 \over 2} \log (r) - {1 - \xi + r^2 \over \sqrt{\lambda (1, r^2, \xi )}} \left \{
{1 \over 2} \log \xi \log \left ( {1 + r^2 - \xi + \sqrt{\lambda (1, r^2, \xi)} \over 1
+ r^2 - \xi - \sqrt{\lambda (1, r^2, \xi )}} \right ) \right . \\
&&- Sp \left ( {2 \sqrt{\lambda (1,r^2, \xi )} \over 1 - r^2 + \xi + \sqrt{\lambda (1,
r^2, \xi )}} \right ) + Sp \left ( {2 \sqrt{\lambda (1, r^2, \xi ) } \over 1 - r^2 - \xi
+ \sqrt{\lambda (1, r^2, \xi )}} \right ) \\
&&- {1 \over 4} \log^2 \left ( {1 - r^2 + \xi - \sqrt{\lambda (1, r^2, \xi )} \over 1 -
r^2 + \xi + \sqrt{\lambda (1, r^2, \xi )}} \right ) + {1 \over 4} \log^2 \left ({ 1 -
r^2 - \xi - \sqrt{\lambda (1, r^2, \xi)} \over 1 - r^2 - \xi + \sqrt{\lambda (1, r^2,
\xi )}} \right ) \\
&&+ {1 \over 2} \ \log^2 \left ( {1 - r^2 + \xi - \sqrt{\lambda (1, r^2, \xi )} \over 2
\xi} \right ) - {1 \over 2} \ \log^2 \left ( {1 - r^2 - \xi - \sqrt{\lambda (1, r^2, \xi
)} \over 2 \xi } \right ) \\
&&\left .  - {1 \over 2} \log^2 \left ( {1 - r^2 + \xi + \sqrt{\lambda (1, r^2, \xi )}
\over 2 \xi} \right ) + {1 \over 2} \ \log^2 \left ( {1 - r^2 - \xi + \sqrt{\lambda (1,
r^2, \xi)} \over 2 \xi} \right ) \right \} \\
&&+ {1 - \xi + 2r^2 \over \sqrt{\lambda (1, r^2, \xi)}} \log \left ( {1 + r^2 - \xi +
\sqrt{\lambda (1, r^2, \xi)} \over 1 + r^2 - \xi - \sqrt{\lambda (1, r^2, \xi)}} \right
) \\
&&- {1 - r^2 \over 2 \xi} \left [ \log (r^2) + {1 - r^2 - \xi \over \sqrt{\lambda (1,
r^2, \xi)}} \log \left ( {1 + r^2 - \xi + \sqrt{\lambda (1, r^2, \xi )} \over 1 + r^2 -
\xi - \sqrt{\lambda (1, r^2, \xi )}} \right ) \right ] \quad . 
  \end{eqnarray*}
\vskip 3 truemm
\noindent And for the Bremsstrahlung :
\begin{eqnarray*}
&&K^{-2,0} = \left [ - 2 + 3 \log \left ( \sqrt{\lambda (1, r^2, \xi )} \right ) - \log
(r) - \log (\xi^{1/2}) \right ] \\
&&- {1 - \xi + r^2 \over \sqrt{\lambda (1, r^2, \xi)}} \log \left ( {1 + r^2 - \xi
+ \sqrt{\lambda  (1, r^2, \xi)} \over 2r} \right ) - {1 + \xi - r^2 \over \sqrt{\lambda
(1, r^2, \xi )}} \log \left ( {1 - r^2 + \xi + \sqrt{\lambda (1, r^2,\xi )} \over 2
\xi^{1/2}} \right ) \\
&& \\
&& \\
&&K^{0,-2} = {1 \over r^2} \left \{ \left [ - 2 + 3 \log \left ( \sqrt{\lambda (1,
r^2, \xi)}\right ) - \log (r) - \log (\xi^{1/2}) \right ] \right . \\
&&\left . - {2(1 - \xi ) \over \sqrt{\lambda (1, r^2, \xi )}} \log \left ( {1 + r^2 - \xi
+ \sqrt{\lambda (1, r^2, \xi)} \over 2r} \right ) - {1 - \xi - r^2 \over \sqrt{\lambda
(1, r^2, \xi )}} \log \left ( {1 - r^2 + \xi + \sqrt{\lambda (1, r^2, \xi )} \over 2
\xi^{1/2}} \right ) \right \} \\
&& \\
&& \\
&&K^{-1,-1} = {1 \over \sqrt{\lambda(1, r^2, \xi)}} \log \left ( {1 + r^2 - \xi +
\sqrt{\lambda (1, r^2, \xi)} \over 2r} \right ) \times \\
&&\left [ - 2 + 4 \log (r) + \log \left ( {1 - r^2 + \xi + \sqrt{\lambda (1, r^2, \xi)}
\over 2 \xi} \right ) + 3 \log \left ( {1 + r^2 - \xi + \sqrt{\lambda (1, r^2, \xi)}
\over 2r} \right ) \right ] \\
&&+ {1 \over \sqrt{\lambda (1, r^2, \xi)}} \left [ \log (r) \log \left ( {1 - r^2 + \xi
+ \sqrt{\lambda (1, r^2, \xi)} \over 2 \xi} \right ) - {\pi^2 \over 3} \right . \\
&&+ 3 \ Sp \left [ \left ( {1 + r^2 - \xi - \sqrt{\lambda (1, r^2, \xi)} \over
2r} \right )^2 \right ] \\
&&\left . - Sp \left [ {1 - r^2 + \xi - \sqrt{\lambda (1, r^2, \xi )} \over 2} \right ] -
Sp \left [ {1 + r^2 - \xi - \sqrt{\lambda (1, r^2, \xi)} \over 2} \right ] \right ] \\
&& \\
&& \\
&&K^{m,n} = {1 \over \pi^2 \sqrt{\lambda (1, r^2, \xi)}} \ I^{m,n} \qquad \hbox{for
$(m,n) = (-1,0), (0,-1)$ }\\ && \\
&& \\
&&K^{m,n} = {1 \over \pi^2 \ m_b^2 \sqrt{\lambda (1, r^2, \xi)}} \ I^{m,n} \qquad \hbox{for $(m,n) = (-1,1), (1,-1)$}
  \end{eqnarray*}

 \end{document}